\documentclass[10pt, conference]{IEEEtran}
\usepackage{amsmath}
\usepackage{amssymb}
\usepackage{amsfonts}
\usepackage{amscd}
\usepackage{mathrsfs}
\usepackage{mdframed}
\usepackage{graphicx}
\usepackage{color}
\usepackage{url}
\usepackage{bm}
\usepackage{algorithm}
\usepackage{algorithmic}

\begin{document}

\title{Insider-Attacks on Physical-Layer Group Secret-Key Generation in Wireless Networks}

\author{J. Harshan$^{\dagger}$, Sang-Yoon Chang$^{\star}$, Yih-Chun Hu$^{\ast}$ $^{\dagger}$\\
$^{\dagger}$Advanced Digital Sciences Center, Singapore, $^{\star}$University of Colorado, Colorado Springs, USA,\\ $^{*}$University of Illinois Urbana-Champaign, USA\\
Email: harshan.j@adsc.com.sg, schang2@uccs.edu, yihchun@illinois.edu\\}


%

\maketitle

%
%

\begin{abstract}
Physical-layer group secret-key (GSK) generation is an effective way of generating secret keys in wireless networks, wherein the nodes exploit inherent randomness in the wireless channels to generate group keys, which are subsequently applied to secure messages while broadcasting, relaying, and other network-level communications. While existing GSK protocols focus on securing the common source of randomness from external eavesdroppers, they assume that the legitimate nodes of the group are trusted. In this paper, we address insider attacks from the legitimate participants of the wireless network during the key generation process. Instead of addressing conspicuous attacks such as switching-off communication, injecting noise, or denying consensus on group keys, we introduce stealth attacks that can go undetected against state-of-the-art GSK schemes. We propose two forms of attacks, namely: (i) different-key attacks, wherein an insider attempts to generate different keys at different nodes, especially across nodes that are out of range so that they fail to recover group messages despite possessing the group key, and (ii) low-rate key attacks, wherein an insider alters the common source of randomness so as to reduce the key-rate. We also discuss various detection techniques, which are based on detecting anomalies and inconsistencies on the channel measurements at the legitimate nodes. Through simulations we show that GSK generation schemes are vulnerable to insider-threats, especially on topologies that cannot support additional secure links between neighbouring nodes to verify the attacks.
\end{abstract}

%
%
\section{Introduction} 
\label{sec:intro}

In wireless ad hoc networks the participating nodes typically engage in group communication such as broadcasting, multiple-access, or relaying operations, where more than two nodes may be simultaneously involved. These sophisticated protocols are motivated by mission-critical goals such as achieving high-throughput, or supporting resource-constrained nodes (such as bandwidth and power) in the network. In order to secure group communication from external eavesdroppers, the nodes use group secret keys (GSKs) to exchange confidential data, and are also involved in key management tasks such as key generation, distribution and maintenance \cite{MNP}. While strengthening the security of key management is an evolving research field, researchers are concurrently focussing on methods to dynamically generate secret keys by exploiting randomness available in the environment. The purpose of dynamic key generation is either to use it as an independent recipe of generating keys, or to use it to update the existing keys. Dynamic key generation techniques are well known, and have been particularly well studied in the context of wireless two-user physical-layer key generation (PKG) \cite{MMYR}. In PKG, the two participants observe a common source of randomness to generate a common key without leaking information to an external passive eavesdropper. Although, the sources of randomness are innumerable, easily accessible ones are the variations in the wireless channel strengths, such as the received signal strength indicator (RSSI) values, complex baseband signal samples, and the channel estimates. Other than the concept of generating secret-keys for two nodes that are within radio frequency (RF) range, some works \cite{ZHL}-\cite{TLQ2} have also addressed using third-party relay(s) to assist the secret-key generation between nodes that are out of range. 

A generalization of relay-assisted secret-key sharing is the concept of wireless physical-layer group secret key (GSK) generation, wherein a group of nodes in the network generate a secret key by observing a common source of randomness \cite{YeR}, \cite{XCDDL}, \cite{TLQ1}. Unlike two-user wireless PKG, in GSK generation, nodes in the group exchange additional messages apart from the pilots so that all the nodes witness the common source of randomness. GSK generation is applicable in ad hoc wireless networks that operate in mobile environments, e.g., MANETs and VANETs \cite{WLF}.
%
Existing works on GSK have focused on developing protocols for exchanging messages so as to (i) reduce the latency, and (ii) to secure the generated keys from an external eavesdropper \cite{YeR}, \cite{XCDDL}, \cite{TLQ1}. Also, the models assume trusted group members and secure the communication against passive eavesdroppers and active attacks from external nodes. In this work, we deviate at this point, to not to discard the possibility of the presence of insiders that involve in active attacks, that disturb the key generation process. In particular, we incorporate the possibility that some legitimate nodes might be compromised (for e.g., as in \cite{LCC}), and the attackers might enter the system as insiders before the commencement of key generation.

Our specific contributions are: 

\begin{itemize}
\item We introduce a novel threat model with insider-compromise, that prior works have not considered. Some crude attacks in our model are to stop relaying signals in the key-generation process, or to inject noise into the system, or to prevent the neighbouring nodes in achieving consensus on the group secret key. We, however, consider more sophisticated attacks that enable the attacker to manipulate the key generation process and yet go undetected. More specifically, we have proposed the following two types of attacks: (i) different-key attacks, wherein an insider attempts to force different realizations of key at different nodes, especially in a network when the legitimate nodes do not have private channels to verify the attack (for e.g., in chain-topology based networks \cite{LYWC}), and (ii) low-rate key attacks, wherein an insider attempts to slow down the channel variations over time, in order to reduce the \emph{secret-key rate}, which is a measure of how fast the generated keys can be updated (Section \ref{sec:threat_model}).
\item To detect the above attacks, one can envision a scheme that uses a separate communication channel built on neighbour discovery and explicit pair-wise communication with the legitimate neighbours; however, such a straightforward scheme introduces new sets of vulnerabilities and can be expensieve in terms of communication-overhead. Therefore, we apply detection schemes that are based on standard phy-layer techniques such as measuring average power levels and Doppler spread (variations of channel over time) \cite{YTA} to identify anomalies and inconsistencies. We show that the proposed insider-attacks can go undetected on the state-of-the-art GSK generation schemes with a non-negligible rate despite using off-the-shelf channel measurement techniques. This implies that GSK generation schemes are vulnerable to insider-threats, especially on topologies that cannot support additional secure links between neighbouring nodes to verify the attacks. (Section \ref{sec:sim_results}) 
\end{itemize}

\section{System Model for GSK Generation} 
\label{sec:model}

Consider a chain-topology based wireless network of $N$ nodes, denoted by the set $\mathcal{S} = \{1, 2, \ldots, N-1, N\}$, where the nodes want to securely communicate among each other using a GSK. We assume that a GSK is dynamically generated at the nodes by exploiting common randomness in the wireless channels. The wireless channel between Node $i \in \mathcal{S}$ and Node $j = i+1 \in \mathcal{S}$ is denoted by a discrete random process $\{ h_{ij}(l) \in \mathbb{C} ~|~ l = 1, 2, 3, \ldots \}$, where the index $l$ captures the realization of the channels at different time instants. All the channels are assumed quasi-static with identical coherence interval of $T_{c}$ seconds, and thus $\{ h_{ij}(l)\}$ can be viewed as the output of sampling a continuous random process $\{ h_{ij}(t)\}$ at $t = lT_{c}$ for $l = 1, 2, \ldots$. Here, $\{h_{ij}(l)\}$ denotes the set of channel estimates derived from the complex baseband samples at the receivers. We assume that nodes communicate on a single frequency but in time division duplex (TDD) fashion, i.e., no two nodes transmit at the same time. Also, by the virtue of channel reciprocity, we assume that nodes $i$ and $j$ observe the same channel environment between them, i.e., $h_{ij}(l) = h_{ji}(l)$. Further, the network topology is assumed to be fixed over several blocks of coherence intervals.

In the above network model, the nodes intend to generate a GSK by observing an accessible and secure source of common randomness. Unlike two-node physical-layer key generation, which requires a minimum of two rounds of pilot exchanges, the multi-node generalization requires more exchanges; the number of times a node transmits depends on the underlying protocol, and also the chosen source of common randomness. For the chain-topology, the set of all channels for a given $l$ is $\mathcal{H}(l) = \{h_{ij}(l) ~|~ \forall j = i+1\}$. For this case, the common source of randomness is some function $f(\mathcal{H}'(l)) \in \mathbb{C}$ for $\mathcal{H}'(l) \subset \mathcal{H}(l)$. Since each node can broadcast its pilot to a maximum of two nodes, some nodes need to forward their observations so that all the nodes witness $\mathcal{H}'(l)$. 

We assume that the coherence time $T_{c}$ is longer than one round of the GSK generation protocol. A generic protocol for exchanging the observations is given below:
\begin{itemize}
\item For $l \in \{1, 2, \ldots, L \}$
\begin{itemize}
\item For $i \in \{1, 2, \ldots, N \}$
\begin{itemize}
\item Node $i$ broadcasts pilot symbol(s), which are used by the rest of the nodes in its RF range to learn the corresponding channel estimates.
\end{itemize}
End
\item Depending on the target $\mathcal{H}'$, the nodes exchange their acquired set of channel estimates, one after the other until each node learns $f(\mathcal{H}'(l))$.
\end{itemize}
End
\end{itemize}

In practice, GSK generation protocols have to be repeated $L$ times to accumulate the required number of bits for the digital key (say $K$ bits). Therefore, the number $L$ depends on number of bits generated in every round (say $k$ bits). Also, the nodes go through additional rounds of communication over a public channel to arrive at consensus on the generated digital secret. In this work, we only focus on the protocol for sharing observations, as it is fundamental to establishing the common source of randomness. For some examples of physical-layer GSK generation, we refer the readers to \cite{LYWC}, \cite{TLQ1}. 

\subsection{Illustrative Example}
\label{subsec:ie}

\begin{figure}
\begin{center}
\includegraphics[scale=0.37]{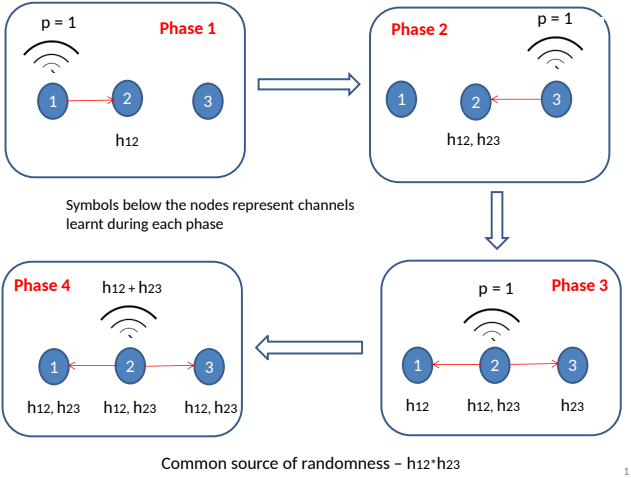}
\caption{\label{fig:gsk_protocol}
A GSK generation protocol between nodes 1, 2 and 3, where $h_{12}$ denotes the channel between nodes 1 and 2, and $h_{23}$ denotes the channel between nodes 2 and 3. The protocol includes 4 phases (read clock-wise from top left), at the end of which all the 3 nodes learn the common source of randomness $h_{12}h_{23}$. 
}
\end{center}
\end{figure}

Consider a small network of three nodes $\mathcal{S} = \{1, 2, 3 \}$ as shown in Fig. \ref{fig:gsk_protocol}, where the nodes 1 and 3 are out of each other's RF range. This implies that nodes 1 and 3 can communicate only through Node 2. We assume that the topology remains fixed for $L$ rounds of the protocol. For a given $l$, the set of underlying channels of the network is $\mathcal{H}(l) = \{h_{12}(l), h_{23}(l)\}$, which is the set of complex channel estimates. For this network, let the common source of randomness be $h_{12}(l)h_{23}(l)$, which implies $\mathcal{H}'(l) = \mathcal{H}(l)$ and $f$ is a product function. To gain insights, we assume that the nodes in the network do not experience additive noise (the noisy case is treated in Section \ref{sec:sim_results}). To achieve the target $h_{12}(l)h_{23}(l)$ at all the nodes, the following protocol is followed during the $l$-th round \cite{SIS}.\footnote{Although the protocol in \cite{SIS} was primarily proposed to facilitate two-node secret generation via a relay, the same technique has been generalized to generate a GSK in mesh networks by \cite{TLQ1}. In this work, we apply the protocol of \cite{SIS} for GSK generation in chain topology.}

\begin{itemize}
\item \textbf{Phase 1}: Node 1 transmits a pilot symbol $p = 1$, which is used by Node 2 to learn $h_{12}(l)$, i.e,
\begin{equation}
\Theta_{1, 2}(l) = h_{12}(l).
\end{equation}
Here $\Theta_{i, j}(l)$ denotes the observation seen by Node $j$ at the end of \textbf{Phase $i$}.
\item \textbf{Phase 2}: Node 3 transmits a pilot symbol $p = 1$, which is used by Node 2 to learn $h_{23}(l)$, i.e,
\begin{equation}
\Theta_{2, 2}(l) = h_{23}(l).
\end{equation}
\item \textbf{Phase 3}: Node 2 transmits a pilot symbol $p = 1$, which is used by Node 1 to learn $h_{12}(l)$, and Node 3 to learn $h_{23}(l)$, i.e.,
\begin{equation}
\Theta_{3, 1}(l) = h_{12}(l), \Theta_{3, 3}(l) = h_{23}(l).
\end{equation}
\item \textbf{Phase 4}: Finally, Node 2 broadcasts the sum $h_{12}(l) + h_{23}(l)$ to the other nodes. Node 1 and Node 3 receive $h_{12}(l)(h_{12}(l) + h_{23}(l))$ and $h_{23}(l)(h_{12}(l) + h_{23}(l))$, respectively, i.e,
\begin{eqnarray}
\Theta_{4, 1}(l) = h_{12}(l)(h_{12}(l) + h_{23}(l)),\\
\Theta_{4, 3}(l) = h_{23}(l)(h_{12}(l) + h_{23}(l)).
\end{eqnarray}
Since nodes 1 and 3 already know one of the components in the linear combination, they cancel the self-interference to retrieve $h_{12}(l)h_{23}(l)$, i.e., 
\begin{equation}
\Theta^{sc}_{4, 1}(l) = h_{12}(l)h_{23}(l), \Theta^{sc}_{4, 3}(l) = h_{12}(l)h_{23}(l).
\end{equation}
Meanwhile, since Node 2 has the realizations of two channels, it computes the product to acquire the common randomness. 
\end{itemize}
The above protocol is depicted in Fig. \ref{fig:gsk_protocol} by dropping the index $l$. In this illustrative example, the order of the first three phases \textbf{Phase 1}, \textbf{Phase 2} and \textbf{Phase 3 } can be permuted in any of 6 possible ways. 

%

%
%
\section{Threat Model}
\label{sec:threat_model}

We study insider-attacks from one of the nodes in the network (specifically the middle node), and leave open the general problem of multiple-insiders or collaborative insider-attacks for future research. We assume that the attacker in the network has the following capabilities: \emph{Assumption 1:} The insider can transmit signals with higher power than the rest of the nodes. \emph{Assumption 2:} The insider has large storage size to store the past channel realizations of the network. \emph{Assumption 3:} The insider can learn the topology of the network. 

We envisage the following two types of attacks from an insider during the key generation process. 

\subsection{Different-Key Attack}
\label{subsec:diff_key_attack}

In GSK generation, the nodes have to securely exchange their observations so that they view correlated source of randomness. A potential threat is the different-key attack, wherein an insider forwards messages to partition the network into two or more subgroups such that nodes across subgroups observe uncorrelated source of randomness. As a result, the nodes across subgroups are forced to generate a different digital-key than the rest of the nodes in other subgroups. If the insider achieves such an attack, then the nodes believe that they have the correct group key. However, during the data communication phase, they will fail to accurately recover the group messages due to mismatched keys. \emph{Assumption 2} and \emph{Assumption 3} may have to be invoked to realize this attack. We exemplify a different-key attack on the protocol of Section \ref{subsec:ie}, by introducing the attack parameters $\{\rho, \rho_{1}, \rho_{2}\}$ such that the special case of $\{\rho = 1, \rho_{1} = h_{12}, \rho_{2} = h_{23}\}$ represents the no-attack case.

\subsubsection{Different-key attack on the chain-topology by Node 2}
\label{subsec:dka_node2}
For the protocol discussed in Section \ref{subsec:ie}, Node 2 gets two chances to sabotage the key-generation process in every round; first time when it transmits the pilot $p = 1$, and the next time when it transmits the sum of two channels $h_{12}(l) + h_{23}(l)$. The most general attack scenario from Node 2 is the following:

\begin{itemize}
\item In \textbf{Phase 3}, Node 2 transmits $p = \rho(l) \in \mathbb{C}$ other than $p = 1$.
\item In \textbf{Phase 4}, Node 2 transmits $\rho_{1}(l) + \rho_{2}(l)$ instead of $h_{12}(l) + h_{23}(l)$, where $\rho_{1}(l)$ and $\rho_{2}(l)$ are chosen based on the attack-objective. 
\end{itemize}

Since $p = \rho(l)$ in \textbf{Phase 3}, the nodes 1 and 3 believe that their channels to Node 2 are $\rho h_{12}(l)$ and $\rho h_{23}(l)$, respectively. Upon completion of \textbf{Phase 4}, the observations at Node 1 and Node 3 after cancelling self-interference are respectively given by, 
\begin{equation*}
\Theta^{sc}_{4, 1}(l) = h_{12}(l)\left(\rho_{1}(l) + \rho_{2}(l)\right) - \rho^2(l) h_{12}(l),
\end{equation*}
\begin{equation*}
\Theta^{sc}_{4, 3}(l) = h_{23}(l)(\rho_{1}(l) + \rho_{2}(l)) - \rho^2(l) h_{23}(l).
\end{equation*}
The objective of the different-key attack is to make sure that $\Theta^{sc}_{4,1}(l)$ and $\Theta^{sc}_{4,3}(l)$ are uncorrelated, which can be achieved by choosing $\rho_{1}(l), \rho_{2}(l)$ statistically independent of $h_{12}(l), h_{23}(l)$.  In this example, \emph{Assumption 2}  has to be invoked in order to store the channels $\{\rho_{1}(l), \rho_{2}(l)\}$, and \emph{Assumption 3} has to be invoked for Node 2 to learn the chain-topology. 

In \eqref{eq:digital_dka} - \eqref{eq:digital_dka3}, we illustrate the effect of the different-key attack on the digital keys, by quantizing the values of $|\Theta^{sc}_{4, 3}|$ and $|\Theta^{sc}_{4, 1}|$ into two levels. Although more sophisticated algorithms exist in literature to generate digital keys \cite{MMYR}, we have used a primitive method to showcase the effect of the proposed attacks on the digital keys.
\begin{eqnarray}
\label{eq:digital_dka}
\mathcal{K}_{\mbox{no attack}} & = & \left[0  ~1 ~ 1 ~ 1 ~ 1 ~ 0 ~ 1 ~ 1 \right],\\
\label{eq:digital_dka2}
\mathcal{K}^{1}_{\mbox{attack}} & = & \left[0 ~ 1 ~ 0 ~ 1 ~  1  ~    0 ~    0  ~   1 \right],\\
\label{eq:digital_dka3}
\mathcal{K}^{3}_{\mbox{attack}} & = & \left[0 ~ 0 ~  1  ~   1  ~   0  ~   1 ~    0  ~   1\right].
\end{eqnarray}

\noindent In the above example, 8-bit digital group keys are generated with (denoted by $\mathcal{K}^{1}_{\mbox{attack}}$ and $\mathcal{K}^{3}_{\mbox{attack}}$ for Node 1 and 3, respectively) and without (denoted by $\mathcal{K}_{\mbox{no attack}}$) the different-key attack by Node 2. With the different-key attack, the keys generated at Node 1 and Node 3 are different.



\subsection{Low-Rate Key Attack}
\label{subsec:low-rate_key_attack}


In physical-layer GSK generation, the number of bits generated from every round of the protocol determines the key-rate. To illustrate further, if $k$ bits of GSK are generated from each round, and the absolute time difference between successive rounds is $\tau$ seconds, then the key-rate is $\frac{k}{\tau}$ bits/sec. This implies that to generate a random bit sequence of length $K$, say $K = 128$ for application under advanced encryption standard, the key generation method requires $\frac{K \tau}{k}$ seconds. Thus, if we wish to update the $K$-bit key frequently, then we should use wireless channels that support high entropy, and/or use protocols that have low-latency. 

With the above discussion on key-rate, a possible insider-attack is to lower the key-rate. This can be accomplished either by reducing the entropy of the channel, i.e., by reducing the number $k$, or by enforcing a slow-varying channel so that the accumulated bits look structured rather than random. \emph{Assumption 1} and \emph{Assumption 2} may have to be invoked to realize this attack. We illustrate a low-rate key attack on the protocol of Section \ref{subsec:ie}. In the next subsection, we introduce the attack parameters $\{\rho, \rho_{1}, \rho_{2}\}$ for the low-rate attack such that the special case of $\rho = \rho_{1} = \rho_{2} = 1$ represents no-attack.

\subsubsection{Low-rate key attack on the chain-topology by Node 2}
\label{subsec:lka_node2}
To execute a low-rate key attack, the fundamental criterion is to enforce $\Theta^{sc}_{4,1}(l)$ and $\Theta^{sc}_{4,3}(l)$ are identical and vary slowly over time. An attack strategy from Node 2 is given below:

\begin{itemize}
\item In \textbf{Phase 3}, instead of $p = 1$, Node 2 transmits $p = \rho(l) \in \mathbb{C}$.
\item In \textbf{Phase 4}, Node 2 transmits $\rho_{1}(l)h_{12}(l) + \rho_{2}(l)h_{23}(l)$ instead of $h_{12}(l) + h_{23}(l)$, where $\rho_{1}(l)$ and $\rho_{2}(l)$ suitably chosen. 
\end{itemize}
Under the above threat model, choosing $\rho_1(l) = \rho_2(l) = \rho^2(l)$ forces 
\begin{equation*}
\Theta^{sc}_{4, 1}(l) = \rho^2(l)h_{12}(l)h_{23}(l) \mbox{ and } \Theta^{sc}_{4, 3}(l) = \rho^2(l)h_{12}(l)h_{23}(l).
\end{equation*}
Although nodes 1 and 3 observe identical source of randomness, their realization has been changed by $\rho^2(l)$. With this attack, the nodes are forced to believe that the individual channels are $\rho(l) h_{12}(l)$ and $\rho(l) h_{23}(l)$, instead of $h_{12}(l)$ and $h_{23}(l)$, respectively. For the above model, the attacker can replace $\Theta^{sc}_{4, 1}(l)$ and $\Theta^{sc}_{4, 3}(l)$ by a slow-varying channel $h'(l)$ by choosing $\rho(l)$ as $\rho(l) = \sqrt{\frac{h'(l)}{h_{12}(l)h_{23}(l)}}.$ Thus, this attack forces the nodes to generate a structured digital key rather than a random one. 

In \eqref{eq:digital_lrka} - \eqref{eq:digital_lrka2}, we illustrate the effect of low-rate key attack on the digital keys, by quantizing the values of $|\Theta^{sc}_{4, 3}|$ and $|\Theta^{sc}_{4, 1}|$ into two levels.
\begin{eqnarray}
\label{eq:digital_lrka}
\mathcal{K}_{\mbox{no attack}} & = & \left[1     ~1    ~ 0 ~    1~     0~     0~     0~   1\right],\\
\label{eq:digital_lrka2}
\mathcal{K}_{\mbox{attack}} & = & \left[1 ~1 ~1 ~1 ~1 ~1 ~1 ~1\right].
\end{eqnarray}
In the above example, 8-bit digital group key is generated with (denoted by $\mathcal{K}_{\mbox{attack}}$) and without (denoted by $\mathcal{K}_{\mbox{no attack}}$) the low-rate key attack by Node 2. With the low-rate key attack, the channel realizations are forced to vary slowly over time, thereby ensuring the quantized values to lie in one of the regions.

\section{Detection Techniques and Simulation Results}
\label{sec:sim_results}

In this section, we discuss possible techniques to detect insider-attacks during the key generation protocol. Detection methods can be broadly classified into two types: (i) direct detection and (ii) indirect detection. In the former class, the nodes communicate among each other to verify the correctness of the generated key by exchanging either the common source of randomness or the generated key (preferably over a private channel to shield leakage from an eavesdropper). In the latter class of methods, the nodes do not explicitly exchange keys or common source of randomness, instead they measure some physical characteristics of their channels to identify anomalies or inconsistencies. For the chain-topology based networks, former class of methods suffer from higher communication-overhead (e.g., nodes need high transmit power to reach the farther nodes), and hence, we focus on the following standard methods in the latter class: (i) Measurement of Doppler spread \cite{YTA} (ii) Measurement of average power levels on the channel estimates.

In the rest of the section, we present a comprehensive analysis on the possibility of detecting an insider-attack by measuring some physical characteristics of wireless channels. In our simulations, we incorporate the existence of additive noise at all the receivers. As a result, the observations made at the nodes are erroneous channel estimates (not ideal as in Section \ref{subsec:ie}). To demonstrate the impact of the detection schemes, we realize the attacks of Section \ref{sec:threat_model} on the protocol discussed in Section \ref{subsec:ie}. 

\subsection{Simulation Setup}

For simulations, all the channels in the network $\{h_{12}(l), h_{23}(l)\}$ are generated using the $1$-st order AR model \cite{BaB} as,
\begin{eqnarray*}
h_{12}(l+1) & = & F h_{12}(l) + n_{12}(l),\\
h_{23}(l+1) & = & F h_{23}(l) + n_{23}(l),
\end{eqnarray*}
for $l > 1$, where $F$ is the Doppler spread of the channel, formally defined as
\begin{equation}
\label{eq:jacobian}
F = E\{h_{ij}(l)h_{ij}(l+1)\} = J_{0}(2 \pi f_{d} \tau),
\end{equation}
with $J_{0}(\cdot)$ denoting the zero-th order Bessel function of first kind, $\tau$ is the absolute time difference (in seconds) between the $l$-th and $(l+1)$-th samples, and $f_{d}$ is the Doppler frequency. The Doppler frequency is given by $f_{d} = f_{c}\frac{v}{c}$, where $v$ is the node velocity, $c$ is the speed of light and $f_{c}$ is the carrier frequency of signal transmission. For the simulations, we use $F = 0.98$. The noise components $n_{12}, n_{23}$ are such that $h_{12}(l), h_{23}(l)$ are independent and distributed as circularly symmetric complex Gaussian noise with zero mean and unit variance, denoted by $\mathcal{CN}(0, 1)$. We assume that the nodes employ the state-of-the-art channel estimation algorithms \cite{BESWB} to estimate the channel gains. 

At the end of \textbf{Phase 3}, the noisy estimates at nodes 1 and 3 are given by, 
\begin{equation}
\label{eq:noise1}
\Theta_{3, 1}(l) = h_{12}(l) + z_{3,1}(l), \Theta_{3, 3}(l) = h_{23}(l) + z_{3,3}(l).
\end{equation}
Similarly, at the end of \textbf{Phase 4}, the estimates before self-interference cancellation are
\begin{eqnarray}
\label{eq:noise2}
\Theta_{4, 1}(l) & = & h_{12}(l)(h_{23}(l) + h_{12}(l)) + z_{4,1}(l) \mbox{ and } \nonumber\\
\Theta_{4, 3}(l) & = & h_{23}(l)(h_{12}(l) + h_{23}(l)) + z_{4,3}(l),
\end{eqnarray}
and after self-interference cancellation are
\begin{eqnarray*}
\Theta^{sc}_{4, 1}(l) & = & \Theta_{4, 1}(l) - (\Theta_{3, 1}(l))^2 \mbox{ and }\\
\Theta^{sc}_{4, 3}(l) & = & \Theta_{4, 3}(l) - (\Theta_{3, 3}(l))^2.
\end{eqnarray*}
In \eqref{eq:noise1} and \eqref{eq:noise2}, the noise components $z_{3,j}(l)$ and $z_{4,j}(l)$, for $j \in \{1, 3\}$ capture the errors in channel estimates, and they are assumed to be distributed as $\mathcal{CN}(0, \alpha_{i, j})$, where $\alpha_{i, j}$ denotes their variance values. In practice, $\alpha_{i, j}$ depends on the signal-to-noise ratio (SNR) of the wireless channel, the number of pilots used to estimate the channel, and also the underlying channel estimation method. For our experiments, we use $\{\alpha_{i, j}\}$ such that the normalized mean square error (NMSE) values during \textbf{Phase 3} and \textbf{Phase 4} are held fixed at some $\eta$, i.e., for $j \in \{1, 3\}$,
\begin{equation*}
\frac{E \{|\Theta_{3, j}(l) - h_{j}(l)|^2\}}{E\{|h_{j}(l)|^2\}} = \frac{\alpha_{3, j}}{E\{|h_{j}(l)|^2\}} = \eta,
\end{equation*}
where $h_{1}(l) = h_{12}(l)$ and $h_{3}(l) = h_{23}(l)$. Similarly, 
\begin{equation*}
\frac{E \{|\Theta_{4, j}(l) - h_{j}(l)(h_{1}(l) + h_{3}(l))|^2\}}{E\{|h_{j}(l)(h_{1}(l) + h_{3}(l))|^2\}}  = \eta.
\end{equation*}

Other than considering noisy estimates, we also study the impact of the attacks against the number of observations, which here refers to the number of rounds of the protocol (denoted by $L$). In the following subsections, we share simulation results obtained while detecting the different-key attack and the low-rate key attacks.


\begin{figure}
\begin{center}
\includegraphics[scale=0.37]{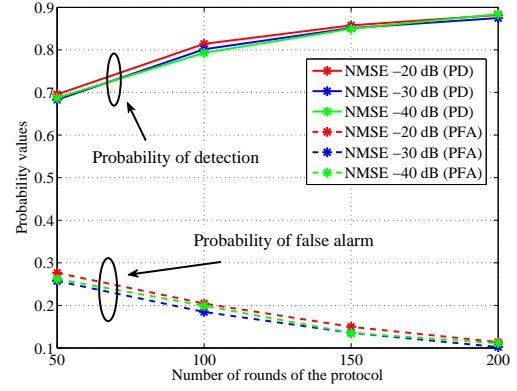}
\vspace{-0.3cm}
\caption{\label{fig:sims_dka_static}
Different-key attack with static values of $\rho_{1}, \rho_{2}$: Probability of detection (PD) and Probability of false alarm (PFA) values as a function of channel estimate quality measured by the normalized mean square error (NMSE) values in dB, and the number of rounds of the protocol. This attack can go undetected even at moderately large values of $L$. 
}
\end{center}
\end{figure}

\begin{figure}
\begin{center}
\includegraphics[scale=0.37]{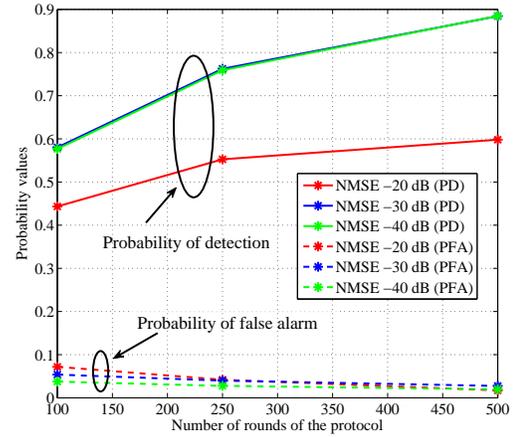}
\vspace{-0.3cm}
\caption{\label{fig:sims_lra}
Low-rate key attack from Node 2: PD and PFA values as a function of NMSE values and the number of rounds of the protocol. This detection technique requires accurate channel estimates (with lower NMSE) and more number of protocol rounds.
}
\end{center}
\end{figure}

\subsection{Different-Key Attacks by Node 2}

We have simulated a different-key attack of Section \ref{subsec:dka_node2} on the protocol in Section \ref{subsec:ie}. We refer to this attack as the static different-key attack, wherein $\rho = 1 ~\forall l$, and $\rho_{1}(l) = h_{12}(1), \rho_{2}(l) = h_{23}(1) ~\forall l$, i.e., $\rho_{1}(l), \rho_{2}(l)$ are fixed throughout several rounds of the protocol. In our experiments, we measure the average power levels on the signals received during \textbf{Phase 3} and \textbf{Phase 4} to calculate the probability of detection (PD) and probability of false alarm (PFA). In particular, the following decision rule was used to detect an attack:
\begin{equation}
\label{1term_inn_prod}
\mbox{DETECT} = \left\{ \begin{array}{cccccccccc}
1, & \mbox{ if } \frac{E\{|\Theta_{4,j}(l)|^2\}}{E\{|\Theta^{sc}_{4,j}(l)|^2\}} < 1, \mbox{ for } j \in \{1, 3\}.\\
0, & \mbox{otherwise}.\\
\end{array}
\right.
\end{equation}
where the expectation is empirically computed over $L$ rounds of the protocol. The above decision rule follows due to self-interference cancellation at nodes 1 and 3. The different-key attack would result in significant difference in the power levels before and after interference cancellation due to non-zero residual terms $\rho_{1}h_{12} - (\rho h_{12})^2$ and $\rho_{2}h_{23} - (\rho h_{23})^2$. However, the residual terms get fully cancelled in the no attack case. Based on the above rule, the attack is detected if atleast one of the nodes output the flag $\mbox{DETECT} = 1$. The results on PD and PFA were obtained by repeatedly running the protocol for $L = \{50, 100, 150, 200\}$ against NMSE values of $\eta = \{-20, -30, -40 \}$ in dB. The results are plotted in Fig. \ref{fig:sims_dka_static}, which show that the static attack gets undetected significant number of times. The plots also show that the probability of false alarm is higher for lower number of observations. This trade-off between PD and PFA can be further optimized by treating the above detection problem as a binary hypothesis problem.

\subsection{Low-Rate Key Attack by Node 2}

We have simulated a low-rate key attack of Section \ref{subsec:lka_node2} in order to force a Doppler spread of $F = 0.99$ on the common source of randomness $\{\Theta^{sc}_{4, 1}(l)\}$ and $\{\Theta^{sc}_{4, 3}(l)\}$, by replacing $h_{12}(l)h_{23}(l)$ by $\rho^{2}(l)h_{12}(l)h_{23}(l)$, which otherwise takes the value of $F = 0.96$ (since $\{h_{12}(l)\}$ and $\{h_{23}(l)\}$ have Doppler spread of $F = 0.98$, the product $\{ h_{12}(l)h_{23}(l)\}$ should have Doppler spread of $0.96$). In our experiments, we have gathered Doppler measurements to compute the probability of detection (PD) and probability of false alarm (PFA). The Doppler computations were obtained as

\begin{small}
\begin{eqnarray*}
F_{3, j} = E \left\lbrace \frac{\Theta_{3, j}(l+1)\Theta^{*}_{3, j}(l)}{|\Theta_{3, j}(l)|^2} \right\rbrace, F_{4, j} = E\left\{\frac{\Theta_{4, j}(l+1)\Theta^{*}_{4, j}(l)}{|\Theta_{4, j}(l)|^2} \right\}
\end{eqnarray*}
\end{small}

\noindent where $F_{3, j}$ and $F_{4, j}$ denote the measured Doppler values on $\{\Theta_{3, j}\}$ and $\{\Theta^{sc}_{4, j}\}$, respectively at Node $j$ for $j \in \{1, 3\}$. These values are empirically computed over $L$ rounds of the protocol. The following decision rule was used on the measured Doppler values:
\begin{equation}
\label{1term_inn_prod}
\mbox{DETECT} = \left\{ \begin{array}{cccccccccc}
1, & \mbox{ if } (F_{3, j} < 0.94) ~\& ~(F_{4, j} > 0.96) \\
0, & \mbox{Otherwise}.\\
\end{array}
\right.
\end{equation}
The above decision rule is coined based on the rationale that the Doppler measurements on $\{\Theta^{sc}_{4,1}(l)\}$ and $\{\Theta^{sc}_{4,3}(l)\}$ should be less than (or equal to) that of $\{\Theta_{3,1}(l)\}$ and $\{\Theta_{3,3}(l)\}$, respectively, in the no attack case. This is because the Doppler spread of the product of two independent channels is lower than or equal to that of each channel in the first-order AR model. We noticed that the \emph{criterion} on average power levels cannot be used to detect this attack since the attack parameters make sure that the self-interference is successfully cancelled at both nodes 1 and 3. The PD and PFA results were obtained for the number of protocol rounds $L = \{100, 250, 500\}$ with NMSE values of $\eta = \{-20, -30, -40\}$ in dB.  The corresponding values of PD and PFA are plotted in Fig. \ref{fig:sims_lra}, which shows that the low-rate attack can go undetected unless measured over large values of $L$, along with accurate channel estimate values. The threshold values on the Doppler spread can be subject to optimization to further improve the trade-off between PD ad PFA, which we defer for future research.

\section{Summary and Future Work}
\label{sec:discussion}
We have chosen a simple 3-node wireless network with chain topology to highlight that GSK generation schemes are vulnerable to insider-threats, especially on topologies that cannot support additional secure links to verify the attacks. 
Without delving into the expensive option of establishing secure links, 
it is an interesting direction for future research to explore other \emph{low communication-overhead} methods that can detect attacks at high success rate by keeping the false positive rates to low values. 

\section*{Acknowledgements}

This work was supported by the research grant for the Human-Centered Cyber-physical Systems Programme at the Advanced Digital Sciences Center from Singapore's Agency for Science, Technology and Research (A*STAR).

%
%

\end{document}